\documentclass[useAMS,usenatbib]{mn2e}

\usepackage{epsfig}

\title[]
{Planetary embryos and planetesimals 
residing in thin debris disks}
\author[]
{
Alice C. Quillen$^{1,3}$,  
Alessandro Morbidelli$^2$,   \&
Alex Moore$^1$ \\
$^1$Department of Physics and Astronomy, University of Rochester, 
Rochester, NY 14627, USA;
aquillen@pas.rochester.edu; \\
$^2$Observatoire de la C\^ote d'Azur, BP 4229, 06304 Nice Cedex 4, France \\
$^3$Visitor, Observatoire de la C\^ote d'Azur \\ 
aquillen@pas.rochester.edu,
morby@obs-nice.fr, 
amoore6@mail.rochester.edu
}

\begin{document}
\label{firstpage}

\maketitle

\begin{abstract}
We consider constraints on the planetesimal population residing
in the disks of AU Microscopii, Beta Pictoris and Fomalhaut 
taking into account their observed thicknesses and
normal disk opacities.
We estimate that bodies of radius 5, 180 and 70 km 
are responsible for initiating the collisional cascade accounting
for the dust production for AU-Mic, Beta-Pic
and Fomalhaut's disks, respectively, at break radii from
the star where their surface brightness profiles change slope.   
Larger bodies, of radius 1000km and with surface density
of order $0.01$g~cm$^{-2}$,
are required to explain the thickness of these disks
assuming that they are heated by gravitational stirring.
A comparison between the densities of the two
sizes suggests
the size distribution in the largest bodies is flatter than
that observed in the Kuiper belt.
AU~Mic's disk requires the shallowest size distribution 
for bodies with radius greater than 10km suggesting that 
the disk contains planetary embryos experiencing a stage of
runaway growth.

\end{abstract}

\section{Introduction}
Recent visible band images taken with the Advanced Camera for Surveys
on the {\it Hubble Space Telescope} well resolve the vertical scale height
of two edge on debris disks, 
the 12Myr old \citep{barrado99,zuckerman01}
dusty circumstellar disks of the M1Ve star 
AU Microscopii (AU~Mic) and the A5V star $\beta$~Pictoris ($\beta$~Pic)
\citep{krist05,golimowski06}.   
Also resolved is the inner edge of Fomalhaut's eccentric ring
also allowing a measurement of the disk scale height
\citep{kalas05}.
The vertical scale height, $H$, is related to the inclination dispersion
of dust particles and so allows an estimate of
the velocity dispersion of the smallest particles.
The velocity dispersion of planetesimals
sets the energy of inter-particle collisions
and so affects a calculation of the dust production rate 
through a collisional cascade
(e.g., \citealt{kenyon02,wyatt02,DD03,wyatt07}).  
The velocity dispersion 
is also sensitive to the presence of larger bodies in the disk
as gravitational scattering or stirring causes an increase
in the velocity dispersion with time (e.g., \citealt{stewartida00,kenyon01}).
Here by combining observations of 
observed vertical thickness with estimates for the dust
production and gravitational
stirring rates we will place constraints on the underlying planetesimal
population in these disks.
Because of the difficulty in resolving vertical
structure, previous cascade calculations have not
used a velocity dispersion
consistent with that estimated for these disks 
or estimated the role of gravitational stirring.

\section{Scaling across the collisional cascade}

We consider three disks with resolved vertical scale heights.
The properties of these three systems along with
the quantities we estimate from them are listed in Table \ref{tab:tab1}.
For AU Mic and $\beta$~Pic we list properties
at the radius, $r$, from the star where there is break in the surface brightness
profile.  For Fomalhaut, we list properties in the ring edge.
One of the observed quantities is the optical depth, $\bar \tau(\lambda)$, 
at wavelength, $\lambda$, normal to the disk plane. 
Because the absorption or emissivity coefficient of a dust
grain with radius $a$ is reduced for $\lambda > a$,
and there are more dust grains with smaller radii,
we expect the optical depth to be related to the number density
of particles of radius $a \sim \lambda$ (e.g., see discussion
in section 4 by \citealt{wyatt02}).
As we only detect the dust particles
in scattered light or in thermal emission,  we use
scaling arguments to estimate the number of larger
bodies residing in the disk.   

Another observed quantity
is the disk thickness that we describe
in terms of a scale height $H$ that here
is a half width.  The disk aspect ratio is the scale height
divided by radius; $h \equiv H/r$.
A population of low inclination orbits has 
$\langle z^2\rangle \approx {r^2 \langle i^2 \rangle \over 2}$,  
so $\bar i \sim \sqrt{2} h $.
Here $\bar i  = \sqrt{ \langle i^2 \rangle}$ and $\langle i^2 \rangle$ is the   
inclination dispersion.
Subsequently we also refer to $\bar e = \sqrt{\langle e^2 \rangle}$ where
$\langle e^2 \rangle$ is the eccentricity dispersion.
We assume a Rayleigh distribution 
of particle inclinations and eccentricities.

We review how the dust opacity and disk thickness can be used to estimate
the planetesimal size distribution. 
Dust production in a destructive collisional cascade can in its simplest
form be studied with a power law size distribution. 
The single power law form for the size distribution is 
in part based on the simplest
assumption that the specific energy 
(kinetic energy per unit mass), $Q_D^*$, required to catastrophically 
disrupt a body is a fixed number independent
of body radius; (often $ 2\times 10^6$erg~g$^{-1}$ for icy bodies is used
based on the estimates by \citealt{kenyon99}).  
The number of particles with radius $a$ 
in a logarithmic bin of size $d \ln a$ 
is predicted to be
\begin{equation}
   {d N \over d \ln a} \equiv N(a) \propto a^{1-q}  
\end{equation}
Using a logarithmic bin gives the same scaling with $a$ as
a cumulative distribution $N_{>a}$ (see appendix A by \citealt{obrien05}).
In an infinite destructive self-similar collisional cascade,
the exponent is predicted to be 
$q = 3.5$ \citep{dohnanyi,tanaka96,davis97,kenyon02}.
The main asteroid belt, if fit with a single
power law, has a lower exponent of $q \sim 2.3$ \citep{ivezic01}. 
It is collisionally evolved but deviates from $q=3.5$
because of additional removal mechanisms 
(e.g., Yarkovsky drift and resonances)
and because the material properties 
depend non trivially on size \citep{obrien05}.  
In contrast the larger bodies in the Kuiper belt are consistent with
$q \sim 5$ \citep{bernstein04}. 
Because of their low number these do not collide
often enough to be part of an ongoing destructive collisional cascade. 
The high exponent probably reflects conditions during the early
solar system when planetesimals were growing as well as colliding
(e.g., \citealt{wetherill93,kokubo96}).

The number of objects of radius $a$ can be estimated 
from another of radius $a_d$ using
the scaling relation 
\begin{equation}
N (a)  = N_d \left({a \over a_d }\right)^{1-q}.
\label{eqn:Na}
\end{equation}
This relates the number of larger particles
to the smallest and so observable particles.
Estimates of the number of dust particles, $N_d$, as a function
of their radius, $a_d$, can be made from 
studies of optical, infrared and submillimeter observations.
It must be kept in mind that because
of the uncertainty in the exponent $q$, 
it is difficult to be accurate
when extrapolating over orders of magnitude in the
size distribution.

The fractional area covered by particles of radius $a$ or 
$\tau(a)$ in a log radial bin can 
be similarly estimated.  Because the opacity depends on the number
per unit area times the cross section area, our assumed power law 
gives for the opacity integrated over a log radial bin 
\begin{equation}
\tau(a) = {d \tau\over d\ln a} = \tau_d \left({ a \over a_d} \right)^{3-q}
\label{eqn:tauq}
\end{equation}
where $\tau_d = \pi a_d^2 s(a_d)$ and $s(a_d)$ is the number of
particles per unit area with radius $a_d$ in a log radial bin.
Likewise the surface mass density 
\begin{equation}
\Sigma(a)  = \Sigma_d \left({ a \over a_d} \right)^{4-q}
\label{eqn:sig_cascade}
\end{equation}
where $\Sigma_d \approx \tau_d \rho_d a_d$.
For $q = 3.5$, most of the disk mass is 
in the largest particles or at the top of the cascade.
Gravitational stirring and dynamical friction 
heating and cooling rates
are proportional to the product of
the surface density time the mass (e.g., 
equations 6.1 and 6.2 by \citealt{stewartida00}), scaling as 
\begin{equation}
\Sigma(a) m (a) = \Sigma_d m_d \left({ a \over a_d} \right)^{7-q},
\label{eqn:sigm}
\end{equation}
where  $\Sigma_d m_d \approx \tau_d \rho_d^2 a_d^4$.
Even when the size distribution is as steep as that
for the large objects in the Kuiper belt ($q \sim 5$)
gravitational stirring is dominated by the largest bodies.

The optical depth
is related to the collision time.  For
a population of identical objects the collision timescale
\begin{equation}
t_{col} \sim (3 \tau \Omega)^{-1}, 
\end{equation}
\citep{hanninen92}
where $\Omega$ is the mean motion (angular rotation
rate for a particle in a circular orbit) at radius $r$.  
Since the collision lifetime is proportional to the inverse
of the optical depth, the timescale for a particle of radius
$a$ to hit another with the same size scale (again
in log radial bins) is
\begin{equation}
t_{col,s} (a ) \approx t_{col,d} \left({a \over a_d} \right)^{q-3}.
\label{eqn:tcolsimple}
\end{equation}
As explored by \citet{DD03,wyatt07}, smaller particles
are capable of dispersing a larger one if the specific energy
of the collision exceeds the critical value.
The collision lifetime is shorter by a factor of $\approx \epsilon^{1-q}$
(Equation 21,22 and associated discussion by \citealt{DD03}),
where $\epsilon^{-1} a$ is the radius of a smaller particle capable of
disrupting one with radius $a$.  The parameter $\epsilon$ is
estimated by considering what energy projectile object can disrupt
the target,
\begin{equation}
\epsilon \sim \left({v_{rel}^2 \over 2 Q_D^*(a)}\right)^{1/3},
\label{eqn:eps}
\end{equation}
(approximating Equation 25 by \citealt{DD03})\footnote{The square root
term in equation 25 by \citet{DD03} should be positive},
where $v_{rel}^2$ is the relative or inter-particle velocity 
dispersion.  We expect the relative velocity
dispersion is twice the particle velocity dispersion
or $v_{rel}^2 \sim 2 u^2$.

We can now estimate the collisional lifetime for particles
in a log radial bin taking into account 
collisions with smaller particles.
After multiplying by Equation \ref{eqn:eps}, 
Equation \ref{eqn:tcolsimple} becomes
\begin{equation}
{t_{col}(a) \over t_{col,d}} \approx \left({a \over a_d} \right)^{q-3}
\left({ u^2 \over    Q_D^*} \right)^{1-q\over 3}.
\label{eqn:tcol}
\end{equation}
For $q=3.5$, the timescale 
$t_{col}(a) \propto a^{0.5}$ consistent with
Equation 23 by \citet{DD03}.
The maximum radius object
that will disrupt during the lifetime of the system
is found by setting $t_{col}(a)$ to the age of
the system, $t_{age}$, and solving Equation \ref{eqn:tcol}
for $a$.  
This estimate was also used by \citet{wyatt02} in their section 5.3.
In other words we define a radius, $a_{top}$ such that
$t_{col}(a_{top}) = t_{age}$ or 
\begin{equation}
a_{top} = a_d 
              \left({ u^2 \over   Q_D^*} \right)^{q-1 \over 3(q-3)}
              \left( t_{age} 3 \tau_d \Omega \right)^{1 \over q-3}.
\label{eqn:atopq}
\end{equation}
For $q=3.5$ this gives
\begin{equation}
a_{top} = a_d  
              \left({ u^2 \over   Q_D^*} \right)^{5 \over 3}
              \left({t_{age}  \over P}\right)^{2}
              \left({ 6 \pi \tau_d }\right)^{2}
\label{eqn:atop}
\end{equation}
where $P$ is the rotation period at radius $r$.
If the disk is hotter or older then a higher surface density
disk that contains more massive bodies
is required to initiate the collisional cascade 
and account for the dust production.

Objects of radius $a_{top}$ are those
likely to be currently initiating the collisional
cascade.    Using Equation \ref{eqn:sig_cascade} with $a_{top}$
we can estimate the total surface density in these massive objects.
As the disk grinds up and is depleted,
more massive but lower number density objects
can enter and generate the cascade. 

\subsection{In relation to observables}

We first relate the disk aspect ratio, $h$,
to the velocity dispersion and the
inclination and eccentricity dispersions.
A population of low inclination orbits has 
$\langle z^2\rangle \approx {r^2 \langle i^2 \rangle \over 2}$,  
so $\bar i \sim \sqrt{2} h $.
An isotropically scattering disk is expected
to have ${\bar i } \sim {\bar e }/2$ (e.g., \citealt{inaba01}).
At low eccentricity, the radial velocity dispersion is
$\langle v_r^2 \rangle  \sim \langle e^2 \rangle v_K^2/2$, 
and the tangential and vertical velocity dispersions are  
$\langle v_\phi^2 \rangle \sim \langle v_z^2 \rangle \sim 
\langle e^2 \rangle  v_K^2/8$, where $v_K$ is the velocity
of a particle in a circular orbit (e.g., see equations C10a,b
by \citealt{wetherill93}).
The total velocity dispersion is the sum
of the three velocity components corresponding to 
$u^2 \sim {3 \over 4}  \langle e^2 \rangle v_K^2$ or 
\begin{equation}
u  \sim  \sqrt{3}~  \bar i v_K \sim \sqrt{6} h v_K.
\end{equation}
These approximations
are consistent with $v_{rel}^2 =(1.25 {\bar e}^2 + {\bar i}^2)v_K^2$
used by previous studies \citep{wyatt02,wetherill93}.

In Equation \ref{eqn:tauq} 
we described the scaling of opacity in a log radial bin.
The normal disk opacity inferred from observations at wavelength $\lambda$,
depends on the disk emissivity or absorption coefficient 
(here denoted $Q$)
\begin{equation}
\bar \tau(\lambda) \approx 
\int_{a_{min}}^{a_{max}} {\tau(a) \over a} Q(\lambda,a) da.
\label{eqn:bartau}
\end{equation}
This is consistent with 
our definition for $\tau(a)$ (Equation \ref{eqn:tauq})
and approximations commonly used in interpreting observed fluxes (e.g.,  
equation 1 by \citealt{backman92} relating
dust opacity to flux and the definition
given in the caption of Fig. 6 by \citealt{pantin97}).
The simplest models for
the absorption or emissivity coefficient of a particle
estimate that these coefficients are 
\begin{equation}
Q(\lambda,a) \approx
\left\{
\begin{array}{ll}
1                                  & {\rm for} ~~ \lambda \le a  \\
\left({\lambda \over a}\right)^{-n} & {\rm for} ~~ \lambda > a 
\end{array}
\right.
\label{eqn:Qabs}
\end{equation}
(e.g., \citealt{backman92,wyatt02}) with $n\sim 1$. 
For $n\sim 1$ and $q\sim -3.5$  
by integrating Equation \ref{eqn:bartau}
we find that $\bar \tau(\lambda) \sim 4 \tau(a = \lambda)$.

More detailed modeling of the absorption coefficients 
(e.g.,  \citealt{pollack94}) shows deviations
from this simplest model  with strong structure
at specific wavelengths such as the $10\mu$m silicate feature.
In addition,
the exponent $q$, describing the dust size distribution, may not be 
well constrained or may not
be the same for small dust particles as for larger ones
(e.g., \citealt{augereau06}) 
or for cm or m sized bodies.  The wavelength at which
the absorption coefficient begins to drop
for equation \ref{eqn:Qabs} may depend on dust composition 
(see discussion in appendix D by \citealt{backman92}).
Multi-wavelength observations are required
to better model the size distribution and composition of the dust.
To take this uncertainty into account we describe
our estimates in 
terms of a factor $f_\tau$, such that 
\begin{equation}
\tau(a=\lambda) = \bar \tau(\lambda)/f_\tau
\label{eqn:ftau}
\end{equation}
that relates the opacity estimated at a wavelength
based on observations
to the size distribution of particles with
radius equal to that wavelength.

An estimate of normal disk opacity at a 
particular radius requires modeling the surface
brightness distribution \citep{golimowski06,krist05,augereau06}. 
Unfortunately,
normal disk opacity estimates are available only
at a few wavelengths for the three disks we are
considering here and not all of
these are based on multi-wavelength models.  
While optical and near-infrared wavelength observations tend to better
resolve the disks, they may not accurately
predict the mm size distribution (e.g., see the discussion
comparing the optical and near-infrared opacities to that
predicted from the submillimeter for AU~Mic by \citealt{augereau06}).
We summarize the existing observed optical depth
measurements for these three disks in Table \ref{tab:tab1} and in
the associated table notes but note
that there is uncertainty in the conversion factor $f_\tau$ between
the measured optical depths and the opacity function
that
we have use here, $\tau(a_d)$, the optical depth integrated
in a log radial bin of size 1 for dust particles of size 
$a_d = \lambda$.
As the opacity of smaller grains is sensitive to the removal
process as well as collisions it is important 
to use observed opacity that is dominated by particles
that are not affected by radiative forces  (e.g., see 
discussion by \citealt{DD03}).

We now convert Equation \ref{eqn:atop} into a form more
easily computed from observables.
The observables are the disk aspect ratio, $h$  and
the normal disk opacity $\bar \tau (\lambda)$ at wavelength $\lambda$.
The size of the objects initiating the collisional cascade
when $q=3.5$
\begin{eqnarray}
a_{top} & \approx &  5.4  {\rm km } 
         \left({\lambda   \over 10\mu{\rm m}}\right) 
         \left({M_* \over M_\odot}\right)^{8 \over 3}
         \left({r   \over 100 {\rm AU}}\right)^{-{14\over 3}}
\nonumber \\
     & &  \times   
        \left({Q_D^* \over 2 \times 10^{6}{\rm erg~g^{-1}}}\right)^{-{5\over 3}}
        \left({t_{age} \over 10^{7} {\rm yr}}\right)^{2} 
        \left({ h \over 0.02} \right)^{10\over 3} 
\nonumber \\
    & & \times
        \left({\bar \tau (\lambda) \over 10^{-2}}\right)^{2}
        \left({ f_\tau  \over 4 }\right)^{-2}
\label{eqn:atopnum}
\end{eqnarray}
Because we have scaled with the inclination or aspect ratio instead
of the collision velocity the exponent of $r$ and
$M_*$ differ from but are consistent with equation 36 by \citet{DD03}.
The relation also differs from previous work \citep{wyatt02,DD03,wyatt07} 
because we have based our estimate on a collision time
scaled from the face on disk opacity 
at a particular radius rather than the
total fraction of starlight re-emitted in the infrared.

Inserting our value for the $a_{top}$ into equation
\ref{eqn:sig_cascade} yields an estimate for 
the total disk density,             
\begin{eqnarray}
\Sigma_(a_{top}) & \approx  & 0.0018 {\rm ~g~cm^{-2}} 
          \left({ \rho_d \over 1 {\rm g~cm^{-3}}}\right)
         \left({M_* \over M_\odot}\right)^{4 \over 3}
         \left({r   \over 100 {\rm AU}}\right)^{-{7\over 3}}
\nonumber \\
 & & \times
         \left({Q_D^* \over 2 \times 10^{6}{\rm erg~g^{-1}}}\right)^{-{5\over 6}}
         \left({t_{age} \over 10^{7} {\rm yr}}\right)
         \left({ h \over 0.02} \right)^{5\over 3}
\nonumber \\
 & & \times
         \left({\lambda   \over 10\mu {\rm m}}\right) 
         \left({\bar \tau (\lambda) \over 10^{-2}}\right)^{2}
         \left({ f_\tau  \over 4 }\right)^{-2}.
\label{eqn:signum}
\end{eqnarray}
We have assumed here that the collision cascade 
started very early in the life of the system, 
however at early stages the inter-particle velocities
were probably not high enough for destructive
collisions \citep{kenyon01,DD03}.  If the timescale
of the destructive cascade were smaller then $a_{top}$
and $\Sigma(a_{top})$ would both be smaller than
the estimates given above.

The product of the density times the mass
for the bodies initiating the cascade $a_{top}$
\begin{eqnarray}
(\Sigma m)(a_{top}) & \approx  & 8.9 \times 10^{15} {\rm g^2~cm^{-2}} 
\nonumber \\
 & & \times
         \left({M_* \over M_\odot}\right)^{28 \over 3}
         \left({r   \over 100 {\rm AU}}\right)^{-{49\over 3}}
\nonumber \\
 & & \times
         \left({Q_D^* \over 2 \times 10^{6}{\rm erg~g^{-1}}}\right)^{-{35 \over 6}}
         \left({t_{age} \over 10^{7} {\rm yr}}\right)^{7 \over 2}
\nonumber \\
 & & \times
         \left({\lambda   \over 10\mu {\rm m}}\right)^4
         \left({\tau_d \over 10^{-3}}\right)^{8}
         \left({ f_\tau  \over 4 }\right)^{-8}  
\nonumber \\
 & & \times
         \left({ h \over 0.02} \right)^{35\over 3}
         \left({ \rho_d \over 1 {\rm g~cm^{-3}}}\right)^2.
\label{eqn:sigmnum}
\end{eqnarray}

\section{Heating the disk with gravitational stirring}

We explore the idea that the observed thickness of the disk
is due to gravitational stirring by bodies of mass, $m_s$, 
surface density, $\Sigma_s$, and size $a_s$.
We define a mass ratio 
$\mu_s \equiv   {m_s \over M_*}$, 
and surface density ratio $\sigma_s \equiv {\Sigma_s r^2 \over M_*}$.
If the disk is in collisional equilibrium then
we expect that $\bar e  \sim 2 \bar i$.

In the dispersion dominated regime,  and assuming that 
the dispersions of the tracer particles exceed those
of the massive particles doing the stirring 
($\bar i > \bar i_s$ and $\bar e > \bar e_s$)
\begin{equation}
{1 \over \Omega} {d \langle i^2 \rangle \over dt}
  \approx {\sigma_s  \mu_s B J_z(\beta)  \beta \over
               \sqrt{\pi} \langle i^2 \rangle }
\label{eqn:gravstir}
\end{equation}
(based on Equation 6.2  by \citealt{stewartida00})
where $\beta = {\bar i \over \bar e} \sim 0.5$ 
(corresponding to Equation 2.11 by \citealt{stewartida00}).
The function described by \citet{stewartida00} 
$J_z(\beta=0.5) \approx 2.0$.
The coefficient $B \sim 2 \ln \Lambda$ and we estimate $\Lambda$
using Equation 2.7 by \cite{stewartida00}
\begin{equation}
\Lambda \approx 3 \mu_s^{-1} {\bar i}^{3}.
\label{eqn:Lam}
\end{equation}
As the coefficient, $B$, only depends logarithmically on $\Lambda$
we can use the scale height estimated from observations
to estimate $\Lambda$ 
and we can solve Equation \ref{eqn:gravstir}
finding that $\bar i \propto t^{-1/4}$, specifically
\begin{equation}
{\bar i}(t) \approx \left( {2 \ln \Lambda \Omega t \sigma_s \mu_s 
                    \over \sqrt{\pi}}  \right)^{1/4}
\end{equation}
The above Equation can be inverted at time $t_{age}$
\begin{equation}
\sigma_s \mu_s \approx {{\bar i}^4 P \over 
                  4 \ln \Lambda \sqrt{\pi} t_{age}},
\end{equation}
where we have set $P$ to be the rotation period at $r$.
In terms of observables this leads to a constraint
on the largest bodies with size $a_s$
\begin{eqnarray}
(\Sigma m) (a_s) & \approx &   
                   2.4 \times 10^{24} {\rm g^2~cm}^{-2} 
                     \left({ h \over 0.02}  \right)^{4}
                     \left({ t_{age} \over 10^7 {\rm yr} } \right)^{-1}
          \nonumber \\ 
 &&   \times         
                     \left({M_* \over M_\odot}\right)^{3\over 2}
                     \left({r \over 100 {\rm AU }}\right)^{-{1 \over 2}}
                     \left({\ln \Lambda \over 12 }\right)^{-1}.
\label{eqn:sigmgrav}
\end{eqnarray}
We note that the constraint on the product of
the surface density times the mass of the largest
bodies is independent of the disk opacity.
In contrast the
estimates for the top of the collisional cascade
(size of object and density) are sensitive to 
the dust opacity. 

\subsection{Connecting the size distributions}

Equation \ref{eqn:atopnum} gives us an estimate for
the size of the bodies at the top of the collisional cascade,
and Equation \ref{eqn:sigmnum} gives us the surface density times mass
in the disk for these bodies.
This product is well
below that needed to account for the disk thickness
with gravitational stirring (Equation \ref{eqn:sigmgrav}).  
To find the size, $a_s$, of
the bodies responsible for the gravitational stirring we
must extend the size distribution beyond $a_{top}$.

Unfortunately,
for bodies with sizes $a> a_{top}$ we can no longer assume
a size distribution consistent with a collisional cascade.
There are few guidelines on what type of power law to
use for bodies greater than 10km.
The only known system that differs significantly
from the size distribution expected from collisional
evolution might
be the largest bodies in the Kuiper belt that have
size distribution with power law
$q \sim 5$ \citep{bernstein04}.
A variety of size distributions might be produced
during the phase of planetesimal growth 
with low values for the exponent $q$ 
at the high mass end implying runaway growth  
(e.g., \citealt{wetherill93,kokubo96,inaba01}).

To place constraints on the size and density of
the largest bodies and exponent of  the size distribution
for these bodies we compare our constraint on the
product of the surface density and mass of the largest
bodies to the surface density and size of the bodies
initiating the collisional cascade.

\begin{figure}
\includegraphics[angle=0,width=3.6in]{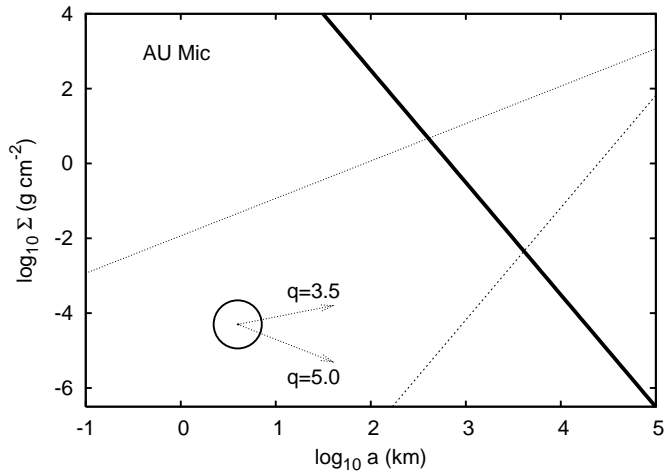}
\caption{
\label{fig:au_connect}
The thick solid line shows the constraint on 
the product of the surface density times mass in the most
massive bodies present for AU~Mic, required to account for the
disk thickness from heating by gravitational stirring.
This is computed using Equation \ref{eqn:sigmgrav} and values
listed in Table \ref{tab:tab1}.
The upper dotted line shows the upper limit on
the surface densities for these
massive bodies set by requiring that they be on averaged
spaced further apart than their mutual hill spheres 
(Equation \ref{eqn:sigh}).
the lower dotted line shows the lower limit on their
surface density set by requiring more than a few bodies of
this mass reside in the disk (Equation \ref{eqn:sigl}).
The large circle is placed at the estimated location of
the top of the collisional cascade (computed using
Equations \ref{eqn:atopnum} and \ref{eqn:sigmnum} and
listed in Table 1).
Arrows are shown with slopes predicted for 
size distributions with $q=3.5$ and $q=5$.
The size distribution must connect the circle
and the segment of the thick solid line that lies
between the two dotted thin lines.
}
\end{figure}

In Figure \ref{fig:au_connect} we plot the constraint
on the product of disk surface density times mass for
AU~Mic.   This constraint corresponds to a surface density
as a function of the radius of a body and is computed
from Equation \ref{eqn:sigmgrav} using values listed
in Table \ref{tab:tab1} and $f_\tau = 4$.
The horizontal axis is log radius instead of log mass so 
the slope of this constraint is -3. The conversion
between mass and radius has been done with a density
of 1~g~cm$^{-3}$.  On this plot we have 
plotted as dotted lines two other constraints on bodies in the disk.
We estimate that the most massive
bodies cannot on average be closer together than their
mutual Hill spheres,
\begin{equation}
\Sigma(m) \la {m \over r_{mH}^2}
\label{eqn:sigh}
\end{equation}
where the mutual Hill radius for two bodies of similar mass
$r_{mH} \equiv r \left({2 m \over 3 M_*}\right)^{1/3}$.
This constraint gives the upper dotted line.
We also require that the number of bodies not be extremely low,
\begin{equation}
\Sigma(m) \ga {10 m \over \pi r^2}.
\label{eqn:sigl}
\end{equation}
This constraint is plotted as the lower dotted line.
The range of densities for the most massive bodies in the disk
must lie on the solid one and between the two dotted ones.
Also plotted on this plot is the estimated density, $\Sigma(a_{top})$,
and radius, $a_{top}$,
of the particles initiating the cascade. 
Arrows are drawn for surface densities $\Sigma(a)$ that 
have size distributions 
with exponents $q=3.5$ and $q=5.0$ and that have 
$\Sigma(a_{top})$.
The circle showing the top of the collisional cascade must
be connected to the thick solid line segment
that lies between the two dotted ones to estimate
the exponent of the size distribution for $a > a_{top}$.

The solid thick line segment between the two thin dotted lines in 
Figure \ref{fig:au_connect} suggests that 1000km bodies reside
in AU~Mic's disk even though the collisional cascade only requires
bodies of radius a few km.   We have checked that
our estimated value of 12 for $\log \Lambda$ is consistent with
the mass of these 1000 km bodies and the disk thicknesses.
(equation \ref{eqn:Lam}).
For $q > 4$ most of the disk mass resides in the most massive bodies. 
Connecting the circle with the line segment requires
a slope shallower than $q = 3.0$.
Most of the disk mass must reside in 
1000 km embryos in AU~Mic's disk to account
for its thickness even though only km sized bodies
are required to account for its dust production.

\begin{figure}
\includegraphics[angle=0,width=3.6in]{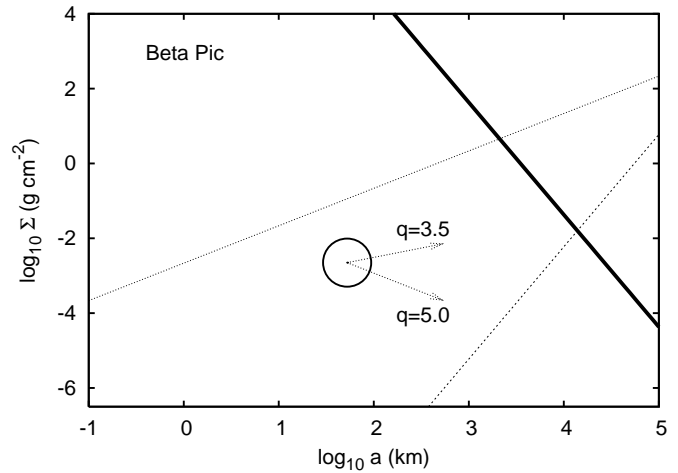}
\caption{
\label{fig:beta_connect}
Similar to Figure \ref{fig:au_connect} except for $\beta$~Pic's disk.
}
\end{figure}

\begin{figure}
\includegraphics[angle=0,width=3.6in]{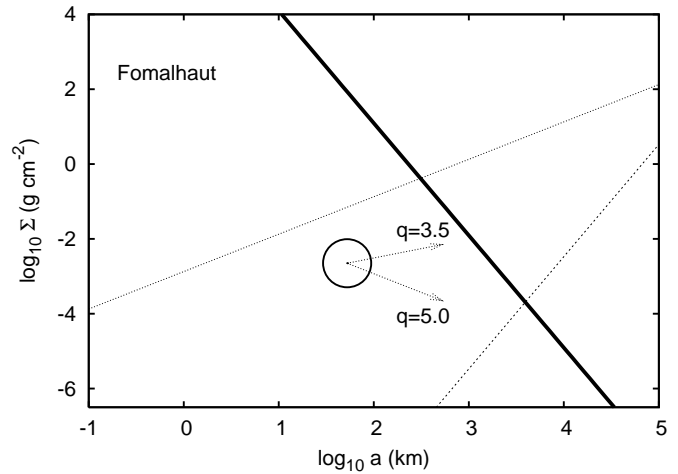}
\caption{
\label{fig:fom_connect}
Similar to Figure \ref{fig:au_connect} except for Fomalhaut's disk.
}
\end{figure}


Figure \ref{fig:beta_connect} and \ref{fig:fom_connect}
are similar to Figure \ref{fig:au_connect} except computed
for $\beta$~Pic's and Fomalhaut's disks also
using parameters listed in Table \ref{tab:tab1}.
We attribute the differences in these figures
primarily to the observed thickness
as $a_{top} \propto h^{10/3}$ (equation \ref{eqn:atopnum}). 
$\beta$-Pic's disk is quite a bit thicker than Fomalhaut's or AU~Mic's so 
its collisional cascade is more efficient and so requires
higher mass progenitors.
Fomalhaut is older allowing a lower density disk to account
for the thickness.   

Gravitational stirring requires similar sized embryos
for the three disks but for  Fomalhaut
the mass and surface density of the bodies is only an
order of magnitude larger than
that predicted from estimating the top of the collisional cascade. 
Nevertheless,
the bodies we infer at the top of the collisional cascade are
not sufficiently dense
and massive to account for the thickness of this disk.

A comparison between the surface densities in the bodies
required to account for the disk thickness
and that predicted at the top of the collision cascade
allows exponents $q \la 3, 3.5$ and  $4.5$ for the three
disks AU~Mic, $\beta$-Pic and Fomalhaut, respectively.
The extremely shallow exponent for AU~Mic at the top end
suggests that the size distribution
deviates from power law form. 
A curve in the size distribution at the high mass end has
been predicted by models and simulations
of planetesimal accretion
when the disk contain embryos in a stage of runaway growth 
\citep{wetherill93,kokubo96,inaba01}.

We have only considered
the effect of gravitational stirring 
in the dispersion dominated regime.
Now that we have an estimate for the masses
of the most massive bodies residing in these disks,
we check this assumption.
Only for $a>1.2\times 10^4$km does a body's Hill
radius approach a scale height $r \bar i$ for an
inclination $\bar  i = 0.01$.
The dispersion dominated
gravitational stirring estimate used in Equation \ref{eqn:gravstir} 
(rather than a sheer dominated one) is therefor reasonable.
Previous work has found that passage through the sheer
dominated regime is comparatively fast
(e.g., \citealt{kenyon01}). 
A better estimate would take into account both regimes, though the
improved constraints on the massive bodies should not significantly
deviate from those estimated here.

\section{Discussion}

We have used estimates of collisional cascades 
(e.g., \citealt{kenyon02,DD03,wyatt07}) to
estimate the size  and surface density of the bodies responsible
for initiating the collisional cascade.
We have done this for 3 debris disks, that of AU Mic,
$\beta$-Pic and Fomalhaut, with resolved
vertical structure estimating
that these bodies have radii of 4, 180, and 70 km, respectively.  
We have estimated these at the radius at which
the surface brightness profile changes slope (also called
the break radius).
The body sizes are a few times 
larger than previous estimates (e.g., \citealt{wyatt02}).  
The differences arise 
because we have based our estimate on a collision time
scaled from the face on disk opacity 
at a particular radius rather than the
total fraction of starlight re-emitted in the infrared and
we have used the observed disk aspect ratio to estimate
the velocity of collisions.

Assuming that the smallest particles are heated solely by
gravitational stirring from the largest ones, the disk
thickness can be used to place a constraint on the product
of the surface density times mass of the largest bodies 
(Equation  \ref{eqn:sigmnum}).
From this we infer that 1000km radius bodies or
planetary embryos are likely
to reside in these three disks.
The large body sizes do not conflict with the lack
of observed gaps in the disks \citep{quillen06,quillen07}
except possibly for the extreme high mass end allowed for
$\beta$-Pic's disk.
A comparison between the surface densities in these bodies
and that predicted at the top of the collision cascade
allows exponents $q \la 3, 3.5, 4.5$ for the three
disks AU~Mic, $\beta$-Pic and Fomalhaut, respectively.
The shallow exponent for AU~Mic at the top end
suggests that this disk
contains embryos in a stage of runaway growth, 
as predicted by simulations \citep{wetherill93,kokubo96,inaba01}.
For all three disks we infer that most of the disk mass 
is likely to reside
in embryos and estimate that the surface densities
are of order $10^{-2}$g~cm$^{-2}$.

A number of simplifying assumptions went into estimating
the properties of the top of the cascade.
We assumed only a single power law form for the size distribution, 
however, the 
specific energy for dispersion is predicted to depend
on body size \citep{benz99} so a single power law is probably
not a good assumption.  
The disks may not have been
sufficiently excited for efficient dust production during
the entire lifetime of these systems  
\citep{DD03}. 
A shorter collisional lifetime would lead to a lower 
surface density and size
estimated for the top of the cascade (see Equations \ref{eqn:signum},
\ref{eqn:atopnum}), though taking into
account the dependence of the specific energy on size
in the regime where self-gravity is important
would increase the surface density of larger bodies
and might decrease the size at the top of the cascade.
The sizes at the top of the cascade predicted here are nearing
the threshold for a destructive equal-mass collision at
a velocity estimated from 
the disk thickness, particularly in the case of Fomalhaut
that has a very thin disk but has a large estimated $a_{top}$. 

Our estimate of the gravitational stirring rate neglected
the role of dynamical friction from smaller particles and
the sheer dominated regime.   Both should be taken
into account to improve the estimate of size and number
of the largest bodies residing in these disks.

Better modeling of the dust distribution using multi-wavelength
observations and high angular resolution imaging would significantly
improve constraints on the small radius end of the size distribution.
While we have found normal disk opacity measurements in a
few wavelengths in the
literature, the different wavelength estimates, 
different assumptions for the assumed size distributions
and different procedures for modeling the data
make it difficult to constrain and compare the dust size distributions
and normal disk opacities among the disks.

We have discussed ways to improve the estimates introduced
here. We now discuss possible implications based
on these predictions.
If the size distributions inferred here are common then
longer lifetimes would be predicted for dust production
because the larger bodies (inferred here), entering the cascade later,
contain a reservoir of mass available for dust production
at later times.  The distribution of disk properties
as a function of age can be used
to place constraints on planetesimal growth models
as well as dust production.

We have only considered opacities at particular
radii for these disks.   For AU Mic and $\beta$-Pic we chose
radii at which there is a break (or change
in slope) in the surface brightness profile. 
If the disk aspect ratios
do not strongly vary with radius then Equation \ref{eqn:sigmgrav}
implies that the product of the mass times the surface density
in the largest bodies, $\Sigma m(a_s) \propto r^{-1/2}$ 
is only weakly decaying with radius.
Compare this to $\Sigma(a_{top}) \propto \tau_d^{-2} r^{-7/3}$
and $a_{top} \propto \tau_d^{-2} r^{-14/3}$
predicted via Equations \ref{eqn:atopnum},\ref{eqn:signum}.
Both $\Sigma(a_{top})$ and $a_{top}$ must drop
rapidly with radius.  If disks
are not extremely thin at larger radii then either there
is another source of heating at large radii accounting
for the disk thickness, or
dust particles detected at large radii originate
from inner radii and are either blown out 
or are on highly eccentric orbits \citep{augereau06,strubbe06}. 
A thin and sparse disk will 
not efficiently produce dust as the collisions
are not destructive.  Consequently multi-wavelength 
observations resolving disks as a function of radius 
should be able to test the utility of the estimates
explored here as well as better
probe planetesimal growth and evolution with radius.

\vskip 1.0 truein
---------------

We thank the Observatoire de la C\^ote D'Azur for support, 
a warm welcome and and hospitality during January 2007.
We thank Patrick Michel, Derek Richardson and Hal Levison 
for interesting discussions.
Support for this work was in part
provided by National Science Foundation grants AST-0406823 \&
PHY-0552695,
the National Aeronautics and Space Administration
under Grant No.~NNG04GM12G issued through
the Origins of Solar Systems Program,
and HST-AR-10972 to the Space Telescope Science Institute.

{}

\begin{table*}
\begin{minipage}{120mm}
\caption{Debris Disks with measured thicknesses}
\label{tab:tab1}
\begin{tabular}{@{}llcccc}
\hline
\multicolumn{5}{l}{Stellar and Disk Properties }\\
Row &                   & AU~Mic     & $\beta$~Pic  & Fomalhaut \\ 
1  & $M_*$($M_\odot$)   & 0.59       & 1.75         & 2.0     \\   
2  & Age (Myr)          & 12         & 12           & 200     \\  
3  & $r$(AU)            & 30         & 100          & 133     \\
4  & $h$                & 0.019      & 0.05         & 0.013   \\
5  &$\bar\tau(\lambda,r)$ 
                        & $3\times 10^{-3}$ 
                                     & $5 \times 10^{-3}$    
                                                    &$ 1.6 \times 10^{-3}$\\
6  & $\lambda$ ($\mu$m) & 1          & 10           & 24      \\
\hline
\multicolumn{5}{l}{Estimated Planetesimal Properties }\\
7  & $a_{top}$(km)      & 4          &  180         & 68      \\
8  & $\Sigma(a_{top})$
     (g~cm$^{-2}$)      & 0.00005    & 0.005        & 0.002   \\
9  & $\Sigma m(a_{top})$
     (g$^2$~cm$^{-2}$)  &$10^{14.5}$ &$10^{21.0}$   &$10^{18.8}$ \\
10 & $\Sigma m(a_s)$
     (g$^2$~cm$^{-2}$)  &$10^{24.1}$ &$10^{26.2}$   &$10^{22.7}$ \\
\hline
\end{tabular}
{ \\
By Row.
1) References for the stellar masses: 
  \citet{houdebine94,crifo97,song01}, respectively.
2) References for the ages:
  \citet{barrado99,barrado98}.
3) The radii are chosen to be where there is a break in the surface
brightness profile as described by 
  \citet{krist05,golimowski06,kalas05}, respectively.
4) The aspect ratio $h = H/r$ for $H$ the half width half max 
of the disk at radius $r$.  Aspect ratios   
are taken from the same references as the break radii listed
in row 3.  
5,6) The normal disk opacity $\bar\tau$ at wavelength
$\lambda$ is given.
References for normal disk opacities:
The normal disk opacity for AU~Mic is estimated for
$1\mu$ sized particles from Fig. 6 by \citet{augereau06} based on images 
in the optical and near infrared.
That for $\beta$-Pic is taken from Fig. 6 by \citet{pantin97} based
on mid-infrared spectra.
That for Fomalhaut is from Table 1 by \citet{marsh05} predicted
for a reference wavelength of $24\mu$m based on 350, 160 and 70$\mu$m imaging.
7) The radius of objects initiating the collisional cascade, $a_{top}$
is estimated using Equation \ref{eqn:atopnum}.
8) The surface density $\Sigma(a_{top})$ 
is estimated using Equation \ref{eqn:signum}.
9) The product of the surface density times the mass
$(\Sigma m)(a_{top})$ is estimated for bodies  initiating
the collisional cascade.
10) The product of the surface density times the mass is
estimated using Equation \ref{eqn:sigmnum}
for bodies responsible for thickening the disk.
Computed quantities listed in rows 7-10 
have been done with parameter $f_\tau =4$
(defined in Equation \ref{eqn:ftau}).
}
\end{minipage}
\end{table*}

\end{document}